\documentclass[10pt,twocolumn,letterpaper]{article}
\usepackage[utf8x]{inputenc}
\usepackage{iccv}
\usepackage{times}
\usepackage{epsfig}
\usepackage{graphicx}
\usepackage{amsmath}
\usepackage{amssymb}
\usepackage{titling}
\usepackage{enumitem}
\usepackage[export]{adjustbox}
\usepackage[accsupp]{axessibility}  % Improves PDF readability for those with disabilities.

\usepackage[pagebackref=true,breaklinks=true,letterpaper=true,colorlinks,bookmarks=false]{hyperref}

\iccvfinalcopy % *** Uncomment this line for the final submission

\def\iccvPaperID{****} % *** Enter the ICCV Paper ID here

\ificcvfinal\fi

\begin{document}
%%%%%%%%% TITLE
\title{Eformer: Edge Enhancement based Transformer for Medical Image Denoising}

% \and

\author{
    Achleshwar Luthra\thanks{equal contribution} \\
    \and
    Harsh Sulakhe\footnotemark[1]\\
    \and
    Tanish Mittal\footnotemark[1] \\
    \and
    Abhishek Iyer\\
    \and
    Santosh Yadav\\
    \and
    Birla Institute of Technology and Science, Pilani \\
    % \begin{center}
    \small\texttt{ \{f20180401, f20180186, f20190658, f20181105, santosh.yadav \} @ pilani.bits-pilani.ac.in
    }
    \\
    % \texttt{@ pilani.bits-pilani.ac.in}
    % \end{center}
}
\date{\vspace{-3pt}}

\maketitle
\ificcvfinal\thispagestyle{empty}\fi
%%%%%%%%% ABSTRACT
\begin{abstract}
In this work, we present Eformer - Edge enhancement based transformer, a novel architecture that builds an encoder-decoder network using transformer blocks for medical image denoising. Non-overlapping window-based self-attention is used in the transformer block that reduces computational requirements. This work further incorporates learnable Sobel-Feldman operators to enhance edges in the image and propose an effective way to concatenate them in the intermediate layers of our architecture. The experimental analysis is conducted by comparing deterministic learning and residual learning for the task of medical image denoising. To defend the effectiveness of our approach, our model is evaluated on the AAPM-Mayo Clinic Low-Dose CT Grand Challenge Dataset and achieves state-of-the-art performance, $i.e.$, 43.487 PSNR, 0.0067 RMSE, and 0.9861 SSIM. We believe that our work will encourage more research in transformer-based architectures for medical image denoising using residual learning.  
\end{abstract}

%%%%%%%%% BODY TEXT
\begin{figure*}[!t]
% \begin{center}
\includegraphics[width=\linewidth, center]{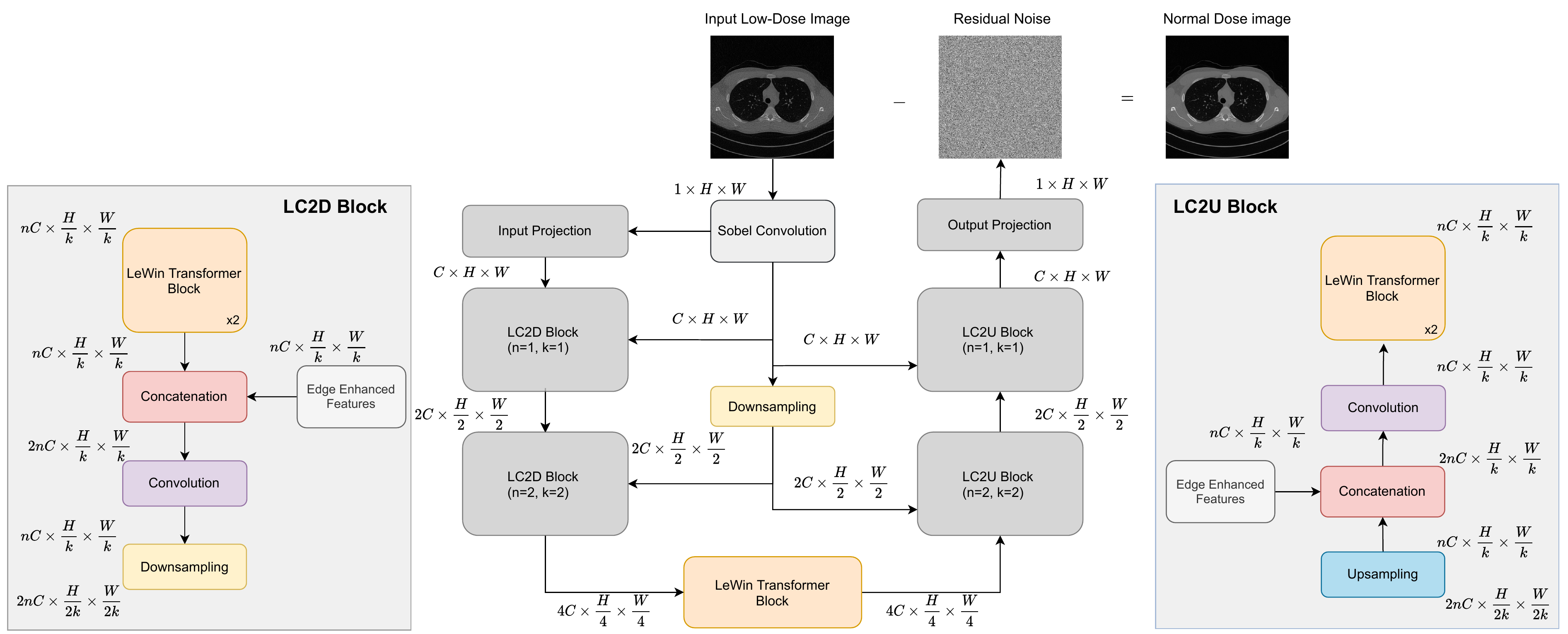}
   \caption{Detailed description of our method. All the steps involved have been explained in \ref{arch}. LC2(D/U) stands for LeWin Transformer, Concatenation block, Convolution block, and Downsampling/Upsampling.}
% \end{center}
\label{fig:method}
\end{figure*}

\vspace{-5pt}
\section{Introduction}

%-------------------------------------------------------------------------
%-------------------------------------------------------------------------
%-------------------------------------------------------------------------

Modern methods for diagnosing medical conditions have been developing rapidly in recent times and a tool of utmost importance is the Computerized Tomography (CT) scan.  It is used often to help diagnose complex bone fractures, tumors, heart disease, emphysema, and more. It works in a method similar to that of the X-Ray scan. A rotating source of X-Ray beams is used to shoot narrow beams through a certain section of your body with a highly sensitive detector being placed opposite to the source which picks up these X Rays and uses a highly advanced mathematical algorithm to create 2D slices of a body part from one full rotation. This process is repeated until a number of slices are created. As helpful as this procedure is in diagnosing, it does have some cause for concern as the patient is exposed to radioactive waves for varying durations. CT scans have been mainly responsible for increasing the radiation received by humans from medical procedures and have even led to medical procedures becoming the second-largest source of radiation after background radiation to affect humans.
Reducing the dose of the X-rays in CT scans is possible but leads to problems such as increased noise, reduction of contrast in edges, corners, and sharp features, and over smoothing of images. We propose a method to help preserve the details and reduce the noise generated from low dose scans so they may become a viable solution in place of high dose scans.

%-------------------------------------------------------------------------
%-------------------------------------------------------------------------
%-------------------------------------------------------------------------

Medical Image Denoising has garnered considerable amount of attention from the computer vision research community. There has been extensive research \cite{Liang_2020, pmid29870364, Chen_2017, pmid31515756, https://doi.org/10.1002/mp.13415, 10.1371/journal.pmed.1002699} in this domain in the recent past. Although these methods have shown excellent results, they implicitly associate denoising with operations on a global scale rather than leveraging the local visual information. We argue that we can benefit from the patch embedding operations that form the basis of a vision transformer \cite{dosovitskiy2021image}. Recently, Vision Transformers (ViT) have shown great success in many computer vision tasks including image restoration \cite{wang2021uformer} but they have not been exploited on medical image datasets. 

%-------------------------------------------------------------------------
%-------------------------------------------------------------------------
%-------------------------------------------------------------------------

To the best of our knowledge, this is the first work that utilizes transformers for medical image denoising. The major contributions of this paper are as follows:

\begin{itemize}
\itemsep0em 
    \item We introduce a novel architecture - Eformer, for edge enhancement based medical image denoising using transformers. We incorporate learnable Sobel filters for edge enhancement which results in improved performance of our overall architecture. We outperform existing state-of-the-art methods and show how transformers can be useful for medical image denoising.
    \item We conduct extensive experimentations on training our network following the residual learning paradigm. To prove the effectiveness of residual learning in image denoising tasks, we also show results using a deterministic approach where our model directly predicts denoised images. In medical image denoising, residual learning clearly outperforms traditional learning approaches where directly predicting denoised images becomes similar to formulating an identity mapping.
\end{itemize}

%-------------------------------------------------------------------------
%-------------------------------------------------------------------------
%-------------------------------------------------------------------------
This paper follows the following structure - in Section \ref{related} we discuss the previous work done in image denoising and the use of transformers in related tasks. In Section \ref{main}, we have explained our approach in a detailed manner. In Section \ref{results}, we compare our results with existing methods which is followed by some conclusive statements and future directions in Section \ref{conclusion}. 

\vspace{-5pt}
%------------------------------------------------------------------------
\section{Related Work}\label{related}

Low-dose CT (LDCT) image denoising is an active research area in medical image denoising due to its valuable clinical usability. Due to the limitations in the amount of data and the consequent low accuracy of conventional approaches \cite{Kaur2018ARO}, data-efficient deep learning approaches have a huge potential in this domain. The pioneering work of Chen $et~al.$ \cite{Chen:17} showed that a simple Convolutional Neural Network (CNN) can be used for suppressing the noise of LDCT images. The models proposed in \cite{Enc-decoder, redcnn, cpce} show that an encoder-decoder network is efficient in medical image denoising. REDCNN \cite{redcnn} combines shortcut connections into the residual encoder-decoder network and CPCE \cite{cpce} uses conveying-paths connections. Fully Convolutional Networks such as \cite{pmid31515756} uses dilated convolutions with different dilation rates whereas \cite{Jifara2019} uses simple convolution layers with residual learning for denoising medical image. GAN based models such as \cite{pmid29870364, https://doi.org/10.1002/mp.13415} use WGAN \cite{arjovsky2017wasserstein} with Wasserstein Distance and Perceptual Loss for image denoising. \\
\vspace{-0pt}
Recently, transformer-based architectures have also achieved huge success in the computer vision domain pioneered by ViT (Vision Transformer) \cite{dosovitskiy2021image}, which successfully utilized transformers for the task of image classification. Since then, many models involving transformers have been proposed that have shown successful results for  many low-level vision tasks including image super-resolution \cite{yang2020learning}, denoising \cite{wang2021uformer}, deraining \cite{chen2021pretrained}, and colorization \cite{kumar2021colorization}. Our work is also inspired by one such denoising transformer, Uformer \cite{wang2021uformer}, which employs non-overlapping window-based self-attention and depth-wise convolution in the feed forward network to efficiently capture local context. We integrate the edge enhancement module \cite{Liang_2020} and a Uformer-like architecture in an efficient novel manner that helps us achieve state-of-the-art results. 

\section{Our Approach}\label{main}

In this section, we provide a detailed description about the components involved in our implementation. 
% Sec \ref{sobel} describes the sobel chttps://www.overleaf.com/project/60ec9094e300f2e6c288b353onvolution used for edge enhancement. In Sec \ref{enc-dec} we describe the transformer based encoder-decoder mechanism. In Sec \ref{down-up} we discuss the techniques that are required at different stages in the architecture, $i.e.$, downsampling and upsampling. In Sec \ref{residue} and Sec \ref{opt}, we explain residual learning and optimization techniques which is followed by overall architecture details in Sec \ref{arch}.  

\subsection{Sobel-Feldman Operator}\label{sobel}
\begin{figure}[!t]
% \begin{center}
% \fbox{\rule{0pt}{2in} \rule{.9\linewidth}{0pt}}
% \end{center}
% \begin{center}
\includegraphics[width=0.75\linewidth, center]{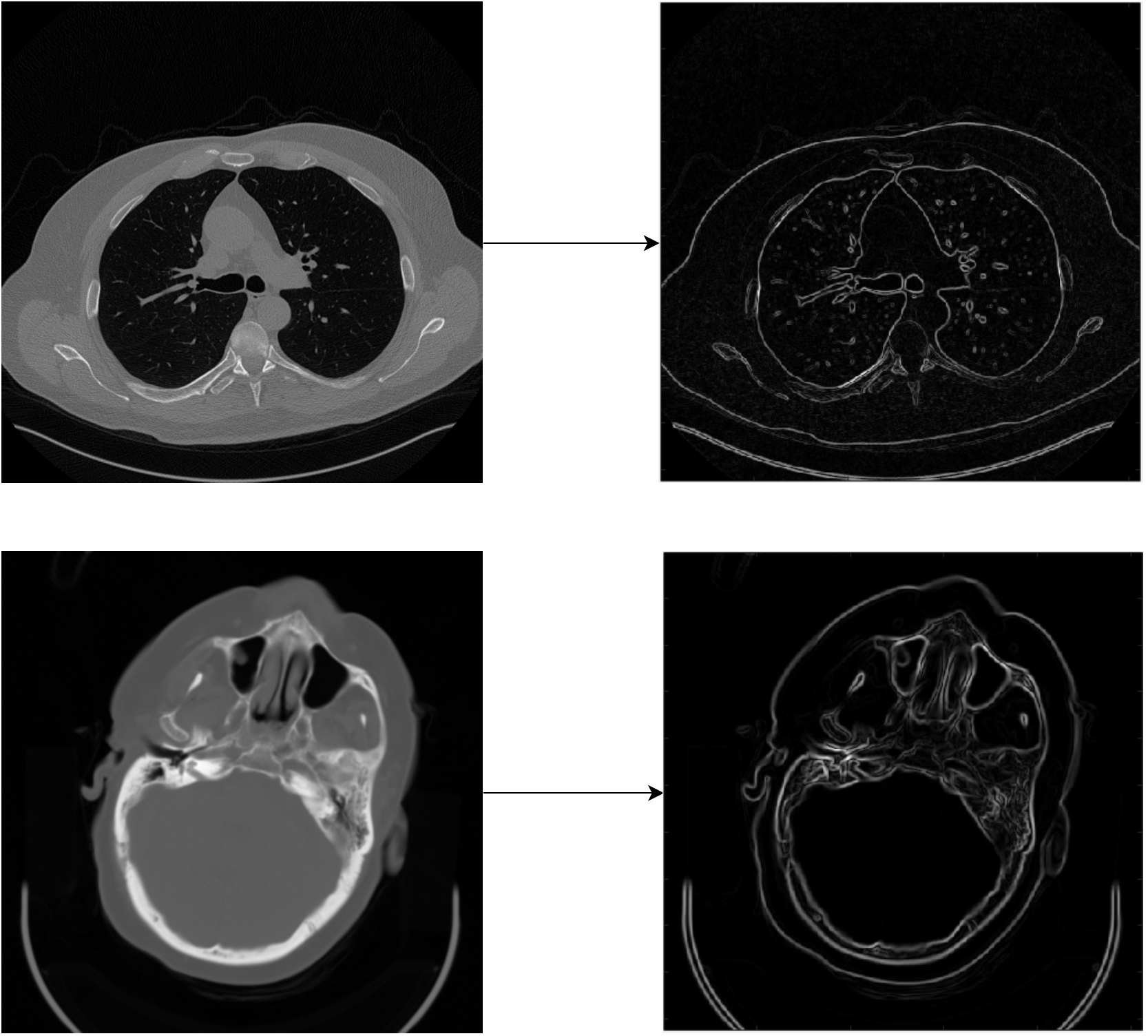}
   \caption{Example of results obtained after convolution of images with Sobel-filter. Input (left) and edge-enhanced images (right).}
% \end{center}
\label{fig:sobel-results}
\end{figure}

% \begin{figure}
% % \begin{center}
% % \fbox{\rule{0pt}{2in} \rule{.9\linewidth}{0pt}}
% % \end{center}
% % \begin{center}
% \includegraphics[width=0.7\linewidth, center]{LaTeX/Images/Sobel-Conv.pdf}
%   \caption{Four different sets of sobel-filters in our implementation.}
% % \end{center}
% \label{fig:sobelfig}
% \end{figure}
Inspired by \cite{Liang_2020}, we use Sobel–Feldman operator \cite{article}, also called Sobel Filter, for our edge enhancement block. Sobel Filter is specifically used in edge detection algorithms as it helps in emphasizing on the edges. Originally the operator had two variations - vertical and horizontal, but we also include diagonal versions similar to \cite{Liang_2020} (See Supplemental Material). Sample results of edge enhanced CT image have been shown in \figurename~\ref{fig:sobel-results}. The set of image feature maps containing edge information are efficiently concatenated with the input projection and other parts of the network (refer to \figurename~\ref{fig:method}). 

\subsection{Transformer based Encoder-Decoder} \label{enc-dec}
 Denoising Autoencoders \cite{redcnn, cpce, Enc-decoder}, Fully Convolutional Networks \cite{Liang_2020, Jifara2019, pmid31515756}, and GANs \cite{pmid29870364, https://doi.org/10.1002/mp.13415} have been successful in the past in the task of medical image denoising, but transformers have not yet been explored for the same, despite their success in other computer vision tasks. Our novel network Eformer is one such step in that direction. We take inspiration from Uformer \cite{wang2021uformer} for this work. At every encoder and decoder stage, convolutional feature maps are passed through a locally-enhanced window (LeWin) transformer block that comprises of a non-overlapping window-based Multi-head Self-Attention (W-MSA) and a Locally-enhanced Feed-Forward Network (LeFF), integrated together (See Supplementary Material) .
\begin{equation}
    \begin{aligned}
   & \mathbf{X}_m^{\prime} = \text{W-MSA}(\text{LN}(\mathbf{X}_{m-1}))+\mathbf{X}_{m-1},  \\
   & \mathbf{X}_m = \text{LeFF}(\text{LN}(\mathbf{X}_m^{\prime}))+\mathbf{X}_m^{\prime}
    \end{aligned}
\end{equation}
here, $\text{LN}$ represents the layer normalization. As shown in \figurename~\ref{fig:method}, the transformer block is applied prior to the LC2D block in each encoding stage and post the LC2U block in each decoding stage, and also serves as the bottleneck layer. 
% $\mathbf{X}_m^{\prime}$ and $\mathbf{X}_m$ are the outputs of the W-MSA and LeFF module respectively. 

% Then, these features use a 4 × 4 convolution with stride 2 for downsampling. A LeWin transformer block is also added in the last of the encoder. Now, features are passed to upsampling layer and LeWin transformer block at each stage in the decoder. For feature reconstruction, stages in the decoder are equal to stages in the encoder.  Similar to the Downsampling layer, Upsampling layer uses 2x2 transposed convolution with stride 2 to match the dimensions to what they were before the downsampling layer in the corresponding encoder stage. After this, up-sampled features and the corresponding features from the encoder through skip-connection are passed to LeWin transformer block\textcolor{red}{[add eqn]}. Features received at the end of all the stages are then passed from a 3 × 3 convolution layer to obtain a residual image R\textcolor{red}{[add eqn]}. 
\begin{table}
\begin{center}
\begin{tabular}{|l|c|c|c|c|}
\hline
Method & MSE & MSP & Adv. & VGG-P \\
\hline\hline
REDCNN \cite{redcnn} & \checkmark & - & - & - \\
WGAN \cite{arjovsky2017wasserstein} & - & - & \checkmark & \checkmark\\
CPCE \cite{cpce} & - & - & \checkmark & \checkmark\\
EDCNN \cite{Liang_2020} & \checkmark & \checkmark & - & -\\
Eformer (ours) & \checkmark & \checkmark & - & -\\
\hline
\end{tabular}
\end{center}
\caption{Comparison between losses used by different methods; MSE - mean squared error, MSP - multi-scale perceptual, Adv. - adversarial, and VGG-P - VGG network based perceptual loss.}
\label{table:loss}
\end{table}

\subsection{Downsampling \& Upsampling} \label{down-up}
Pooling layers are the most common way of downsampling the input image signal in a convolutional network. They work well in image classification tasks as they help in capturing the essential structural details but at the cost of losing finer details which we cannot afford, in our task. Hence we choose strided convolutions in our downsampling layer. More specifically, we use a kernel size of $3\times3$ with stride of $2$ and padding of $1$.
% In medical image denoising, we cannot afford to lose such details 

Upsampling can be thought of as unpooling or reverse of pooling using simple techniques such as Nearest Neighbor. In our network, we use transpose convolutions \cite{dumoulin2018guide}. Transpose convolution reconstructs the spatial dimensions and learns its own parameters just like regular convolutional layers. The issue with transpose convolutions is that they can cause checkerboard artifacts which are not desirable for image denoising. \cite{odena2016deconvolution} states, to avoid uneven overlap, the kernel size should be divisble by the stride. Hence, in our upsampling layer, we use a kernel size of $4\times4$ and a stride of $2$.
% (also called deconvolution). .  

\subsection{Residual Learning}\label{residue}
The goal of residual learning is to implicitly remove the latent clean image in the hidden layers. We input a noisy image $x = y + v$ to our network, here $x$ is the noisy image, in our case the low-dose image, $y$ is the ground truth, and $v$ is the residual noise. Rather than directly outputting the denoised image $\hat{y}$, the proposed Eformer predicts the residual image $\hat{v}$, $i.e.$, difference between the noisy image and the ground truth. According to \cite{he2015deep}, when the original mapping is more like an identity mapping, the residual mapping is much easier to optimize. Discriminative denoising models aim to learn a mapping function of $F(x) = \hat{y}$ whereas we adopt residual formulation to train our network to learn a residual mapping $R(x) = \hat{v}$ and then we obtain $\hat{y} = x - R(x) \implies \hat{y} = x - \hat{v}$. 

\subsection{Optimization}\label{opt}
% As the formula of PSNR is directly related to MSE, The model trained by using MSE as the loss function can get good results. However, Even PSNR cannot truly reflect the visual quality of the output image.  \\
% Therefore, in our model, we are using a weighted loss which combines Perceptual loss, compound loss\cite{zhang2018unreasonable} and MSE Loss.  MSE is the mean of the squared differences between low dose CT image and predicted CT image[add eqs]. Sometimes, 2 images can look very similar to humans but are very different mathematically. So, we used Perceptual loss which is capable of distinguishing images in a similar manner as human. It uses a pretrained neural network to extract high level features. The work of [add ref] has done detailed comparison on how perceptual loss using neural network can produce better results than the conventional loss parameters like MSE. In our model, Perceptual loss uses pretrained VGG model for calculating loss. 

As a part of the optimization process, we employ multiple loss functions to achieve the best possible results. We initially use Mean Squared Error (MSE) which calculates the pixelwise distance between the output and the ground truth image defined as follows.
\begin{equation}
    L_{mse} = \frac{1}{N}\sum_{i=1}^N\Big\|(x_i - R(x_i)) - y_i\Big\|^2
\end{equation}
However, it tends to create unwanted artifacts such as over-smoothness and image blur. To overcome this, we employ both, a ResNet \cite{he2015deep} based Multi-scale Perceptual (MSP) Loss \cite{Liang_2020}. MSP can be described by the following equation 
\begin{equation}
% and a VGG \cite{simonyan2015deep} based perceptual loss (LPIPS) \cite{zhang2018perceptual}
L_{msp} = \frac{1}{NC}\sum_{i=1}^N\sum_{s=1}^C\Big\|\phi_s(x_i - R(x_i),\hat{\theta}) - \phi_{s}(y_i,\hat{\theta})\Big\|^2
\end{equation}
A ResNet-50 backbone was utilized as the feature extractor $\phi$. To be specific, the pooling layers from a ResNet-50 pretrained on the ImageNet dataset \cite{5206848imagenet} were deleted, retaining the convolutional blocks following which the weights ($\hat{\theta}$) were frozen. To calculate perceptual loss, the denoised output $x_i - R(x_i)$, where $R(x_i) = \hat{v_i}$ (as described in Section \ref{residue}) and ground truth $(y_i)$ are passed to the extractor. Following this, feature maps are extracted from four stages of the backbone, as done in \cite{Liang_2020}. 
This perceptual loss, in combination with MSE deals with both per-pixel similarity in addition to overall structural information. Our final objective is as follows,
\begin{equation}
    L_{final} = \lambda_{mse} L_{mse} + \lambda_{msp} L_{msp} 
    % + \lambda_3 L_{lpips}
\end{equation}
where, $\lambda_{mse}$ and $\lambda_{msp}$ are pre-defined constants.

\begin{figure}
% \begin{center}
% \fbox{\rule{0pt}{2in} \rule{.9\linewidth}{0pt}}
% \end{center}
\begin{center}
\includegraphics[width=0.8\linewidth]{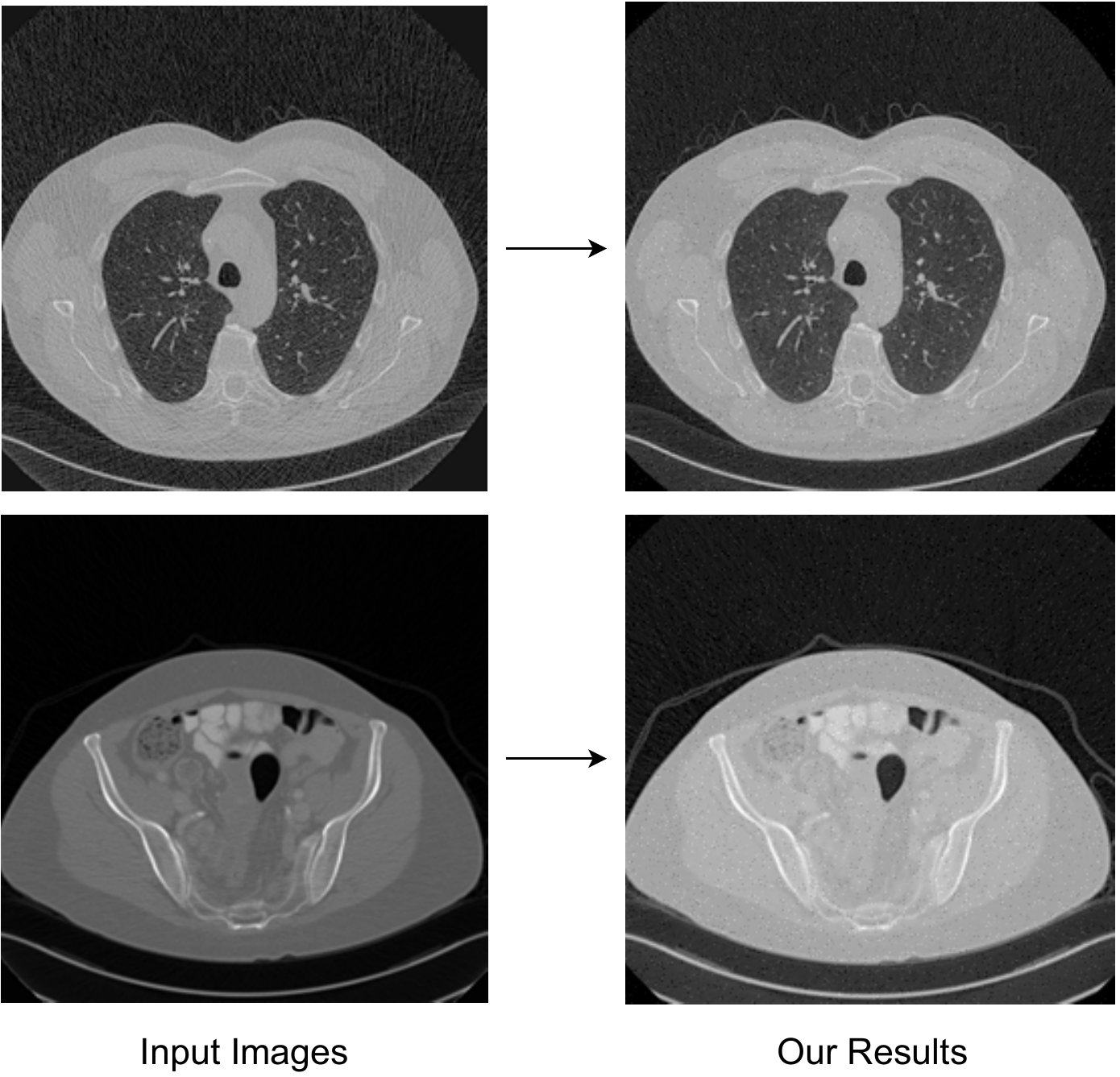}
   \caption{Sample Results on AAPM Dataset \cite{mccollough2017low}. More results have been added in the supplementary material.}
   \label{fig:res}
\end{center}
\end{figure}
% \vspace{-0pt}

\subsection{Overall Network Architecture}\label{arch}
Composing the aforementioned individual modules, our pipeline can be described as follows. An input image $I$ is first passed through a Sobel Filter to produce $S(I)$ followed by a GeLU activation \cite{hendrycks2020gaussian}. As a part of the encoding stages, at each stage, we pass the input through a LeWin transformer block, proceeded by a concatenation with $S(I)$ and consequent convolution operations, similar to \cite{Liang_2020} to produce an encoded feature map. The feature map, along with $S(I)$ is then downsampled using the procedure described in Section \ref{down-up}. Post encoding, at the bottleneck, we pass the encoded feature map to another LeWin Transformer block which is now ready to be decoded by the same number of stages as it was encoded. In each stage of the decoder, post deconvolution, the earlier downsampled $S(I)$ itself are concatenated with the upsampled feature maps which are then passed through a convolutional block. The decoder stage can be viewed as an opposite of the encoder stage, with a shared $S(I)$. The final feature map produced after decoding is then passed through a 'output projection' block to produce the desired residual. This 'output projection' is a convolutional layer, that simply projects the $C$-channel feature map to a $1$-channel grayscale image. In our experiments, we set the depth of the LeWin block, attention heads and number of encoder-decoder stages each to $2$. A concise representation of the architecture can be seen in \figurename~\ref{fig:method} which resembles the alphabet 'E' hence the name Eformer. 

\begin{table}
\begin{center}
\begin{tabular}{|l|c|c|c|}
\hline
\textbf{Method} & \textbf{PSNR $\uparrow$} & \textbf{SSIM $\uparrow$} & \textbf{RMSE $\downarrow$} \\
\hline\hline
REDCNN & 42.3891 & 0.9856 & 0.0076 \\\hline
WGAN & 38.6043 & 0.9647 & 0.0108\\
\hline
CPCE & 40.8209 & 0.9740 & 0.0093\\
\hline
EDCNN & 42.0835 & \textbf{0.9866} & 0.0079\\
\hline\hline
Eformer & 42.2371 & 0.9852 & 0.0077\\
\hline
Eformer-residual & \textbf{43.487} & 0.9861 & \textbf{0.0067}\\
\hline
\end{tabular}
\end{center}
\caption{Comparison with previous methods evaluated on AAPM Dataset \cite{mccollough2017low}.}
\label{table:compare}
\end{table}

\section{Results and Discussions}\label{results}

This subsection highlights the results attained by measuring three different metrics to judge noise reduction  and the quality of the reconstructed low dose CT images. We use the following metrics for the evaluation - Peak Signal to Noise Ratio (PSNR), Structural Similarity (SSIM), and Root Mean Square Error (RMSE). PSNR is targeted at noise reduction and is a measure of the quality of reconstruction. SSIM is a perceptual metric that focuses on the visible structures in an image and is a measure of the visual quality. RMSE keeps track of the absolute pixel to pixel loss between the two images. We compare our results, examples shown in \figurename~\ref{fig:res}, with architectures that share similarities with our model in the sense they are based on a convolutional architecture. As seen in Table \ref{table:loss}, CPCE \cite{cpce}, WGAN \cite{arjovsky2017wasserstein} and EDCNN \cite{Liang_2020} like ours use a combination of commonly used losses to train their model while REDCNN \cite{redcnn} only uses MSE. Table \ref{table:compare} shows that our proposed models, Eformer and Eformer-Residual, outperform the state-of-the-art methods in both the PSNR and MSE metrics, indicating efficient denoising and our comparable performance in SSIM also suggests that the visual quality of the image is high and important details are not lost in the reconstruction. 

\vspace{-5pt}
\section{Conclusion}\label{conclusion}
To conclude, this paper presents a residual learning based image denoising model evaluated in the medical domain. We leverage transformers, and an edge enhancement module to produce high quality denoised images, and achieve state-of-the-art performance using a combination of multi-scale perceptual loss and the traditional MSE loss. We believe our work will encourage the use of transformers in medical image denoising. In the future, we plan to explore the capabilities of our model on a multitude of related tasks.

\section{Acknowledgements}

We want to thank the members of Computer Vision Research Society (CVRS\footnote{https://sites.google.com/view/thecvrs}) for their helpful suggestions and feedback.

{\small
\bibliographystyle{ieee_fullname}
\bibliography{egbib.bib}
}

\newpage
\title{Supplementary Material for Eformer: Edge Enhancement based Transformer for Medical Image Denoising}
\author{
    Achleshwar Luthra\footnotemark[1] \\
    \and
    Harsh Sulakhe\footnotemark[1]\\
    \and
    Tanish Mittal\footnotemark[1] \\
    \and
    Abhishek Iyer\\
    \and
    Santosh Yadav\\
    \and
    Birla Institute of Technology and Science, Pilani \\
    % \begin{center}
    \small\texttt{ \{f20180401, f20180186, f20190658, f20181105, santosh.yadav \}
   @ pilani.bits-pilani.ac.in}
    % \end{center}
}
\maketitle
\section{Dataset Details}
For our research work, we have utilized the AAPM-Mayo Clinic Low-Dose CT Grand Challenge Dataset \cite{mccollough2017low} provided by The Cancer Imaging Archive (TCIA). The dataset contains 3 types of CT scans collected from 140 patients. These 3 types of CT scans are abdomen, chest, and head which are collected from a total of 48, 49, and 42 patients respectively. The data from each patient comprises of low-dose CT scans paired with its corresponding normal-dose CT scans. The low dose CT scans are synthetic CT scans which are generated by poisson noise insertion into the projection data. Poisson noise was inserted to reach a noise level of 25\% of the full dose. Each CT scan is given in DICOM (Digital Imaging and Communications in Medicine) file format. It is a standard format which establishes rules for the exchange of medical images and associated information between different vendors, computers and hospitals. This format meets health information exchange (HIE) standards and HL7 standards for transmission of health-related data. A DICOM file consists of a header and image pixel intensity data. The header contains information regarding the patient demographics, study parameters, $etc.$ stored in seperate 'tags' and image pixel intensity data contains the pixel data of the CT scan which in our case contains pixel data of images of size $512\times512$. In our model, for training, we extracted the image pixel data from a Dicom file to a NumPy array using Pydicom library \footnote{https://pydicom.github.io/} and then, the pixel data in NumPy array is scaled from 0 to 1 to avoid heterogenous spanning of pixel data for different CT scans.

\begin{figure}
% \begin{center}
% \fbox{\rule{0pt}{2in} \rule{.9\linewidth}{0pt}}
% \end{center}
% \begin{center}
\includegraphics[width=\linewidth, center]{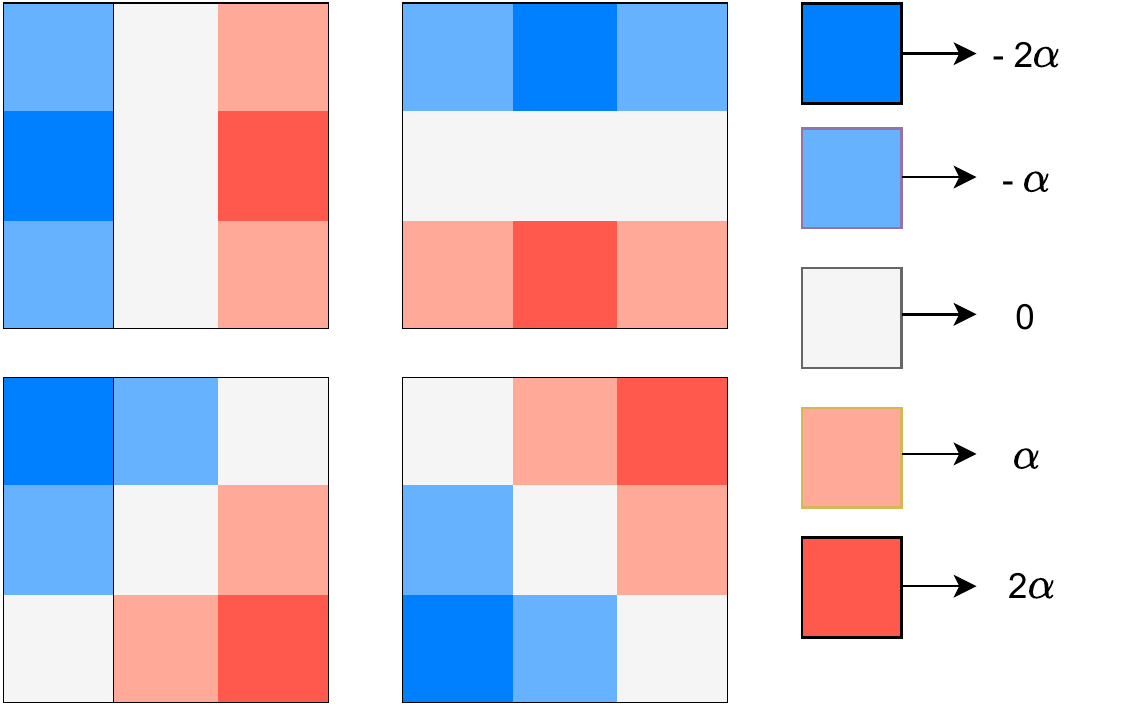}
  \caption{Four different sets of Sobel-filters in our implementation.}
% \end{center}
\label{fig:sobelfig}
\end{figure}

\section{Parameter Details and Network Training}
% parameters
The structure and architecture of the model have been previously described in Section 3.6 and Figure 1 of the main text. We use the Pytorch framework \cite{paszke2017automatic} to run our experiments. The convolutional layers are initialized using the default scheme except the Sobel convolutional block. We enforce the filter parameters to follow the pattern as shown in \figurename~\ref{fig:sobelfig} where $\alpha$ is a learnable parameter. All our experiments were run on a 16GB NVIDIA TESLA P100 GPU. The model was trained with an ADAM \cite{Adam} optimizer using a learning rate of $0.00002$ and default parameters. The model was trained using an input size of $128\times128 $ pixels by resizing each image from its original size of $512\times512$ pixels. The results obtained have been shown in \figurename~\ref{fig:results_supmat}

\begin{figure}
% \begin{center}
% \fbox{\rule{0pt}{2in} \rule{.9\linewidth}{0pt}}
% \end{center}
% \begin{center}
\includegraphics[width=0.55\linewidth, height=9cm, center]{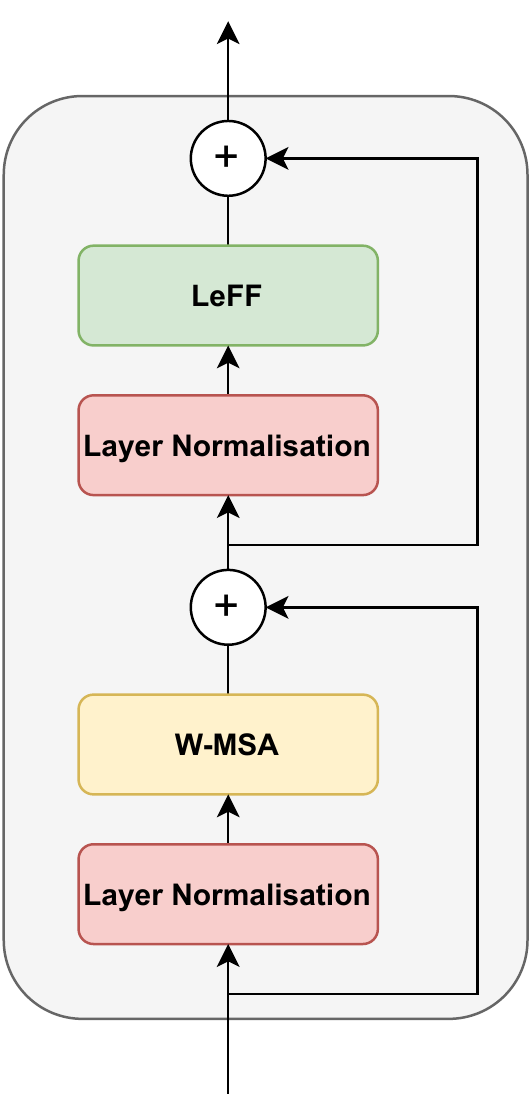}
  \caption{LeWin Transformer Block}
% \end{center}
\label{fig:lewin}
\end{figure}

\begin{figure}
% \begin{center}
% \fbox{\rule{0pt}{2in} \rule{.9\linewidth}{0pt}}
% \end{center}
% \begin{center}
\includegraphics[width=1\linewidth,height=12cm, center]{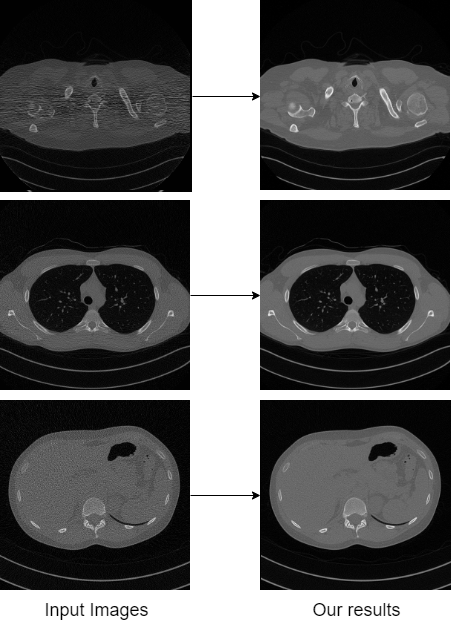}
  \caption{Results}
% \end{center}
\label{fig:results_supmat}
\end{figure}

\section{LeWin Transformer}
To make our submission self-containing, we have provided architecture details of the LeWin transformer block \cite{wang2021uformer} in the supplementary material. LeWin transformer block (\figurename~\ref{fig:lewin}) contains 2 core designs which are described below. First, \textbf{non-overlapping Window-based Multi-head Self-Attention (W-MSA)}, which works on low-resolution feature maps and is sufficient to learn long-range dependencies. Second, \textbf{Locally-enhanced Feed-Forward Network (LeFF)}, which integrates a convolution operator with a traditonal feed-forward network and is vital in learning local context. In LeFF, the image patches are first passed through linear projection layers followed by $3\times3$ depth-wise convolutional layers. Further the patch features are flattened and finally passed to another linear layer to match the dimension of input channels. The structure of the LeWin Transformer Block is pictorially represented in \figurename~\ref{fig:lewin}. Corresponding equations are as follows.
\begin{equation}
    \begin{aligned}
   & \mathbf{X}_m^{\prime} = \text{W-MSA}(\text{LN}(\mathbf{X}_{m-1}))+\mathbf{X}_{m-1},  \\
    \end{aligned}
\end{equation}

\begin{equation}
     \mathbf{X}_m = \text{LeFF}(\text{LN}(\mathbf{X}_m^{\prime}))+\mathbf{X}_m^{\prime}
\end{equation}
Here $\mathbf{X}_m^{\prime}$ and $\mathbf{X}_m$ are the outputs of the W-MSA module and LeFF module respectively, $\text{LN}$ represents layer normalization. In the W-MSA module, the given 2D feature map $X\in \mathbb{R} ^ {C\times H\times W}$ is split into $N$ non-overlapping windows with window size $ M\times M$. Following this, self-attention is performed on the flattened features of each window $X^i \in \mathbb{R}^{M^2 \times C}$. Suppose the head number is $j$ and the head dimension is $d_j = C/j$. Then, consequent computations are,
\begin{equation}
    \mathbf{X} = \{\mathbf{X}^1, \mathbf{X}^2, \dots, \mathbf{X}^N\}, ~~ N = HW/M^2
\end{equation}
\begin{equation}
    \begin{aligned}
       \mathbf{Y}^i_j = \text{Attention}(\mathbf{X}^i\mathbf{W}_j^Q, \mathbf{X}^i\mathbf{W}_j^K, \mathbf{X}^i\mathbf{W}_j^V)\\
    i = 1 \dots N
    \end{aligned}
\end{equation}

\begin{equation}
    \mathbf{\hat{X}}_j =\{\mathbf{Y}^1_j, \mathbf{Y}^2_j, \dots, \mathbf{Y}^M_j\}
\end{equation}
$\mathbf{\hat{X}}_j$ denotes the output for the $j$-th head. Now, output for all the heads can be concatenated and then linearly projected to get the final results. We formulate attention calculation in the same manner as done in \cite{wang2021uformer}.

% {\small
% \bibliographystyle{ieee_fullname}
% \bibliography{supbib.bib}
% }

\end{document}